\documentstyle[preprint,aps,amsfonts]{revtex}
\tighten
\def\vec#1{\bbox{#1}}
\def\i{{\mathrm i}}
\begin{document}
\title{Quantum Fields and Dissipation}
\author{Peter A.Henning,\thanks{
Address after June 1, 1996: Institut f\"ur Kernphysik,
Technische Hochschule Darmstadt, Schlo\ss gartenstr. 9, 
D-64289 Darmstadt}}
\address{Theoretical Physics,
        Gesellschaft f\"ur Schwerionenforschung GSI\\
        Planckstra\ss e 1, D-64291 Darmstadt, Germany\\
        P.Henning\makeatletter @\makeatother gsi.de;
  http://www.gsi.de/\~{}phenning/henning.html}
\date{GSI-Preprint 96-23; May 30, 1996\\
Contribution to {\em Physics Essays}}
\maketitle
\begin{abstract}
The description of thermal or non-equilibrium systems necessitates
a quantum field theory which differs from the usual approach in two
aspects: 1.The Hilbert space is doubled; 2.Stable quasi-particles 
do not exist in interacting systems. A mini-review of these two
aspects is given from a practical viewpoint including two
applications. For thermal states it is shown how infrared divergences
occuring in perturbative quasi-particle theories are avoided, whereas
for non-equilibrium states a memory effect is shown to arise in the
thermalization.
\end{abstract}
\section{Introduction}
When the first decades of our century saw the rapid ascent of
quantum theory, physicists were troubled by the question:
{\em How can dissipation and irreversibility arise in the macroscopic 
nature, although the microscopic equations governing the world are 
time-reversal invariant ?\/} The only place where dissipation was visible 
in quantum mechanics at this time was the {\em decay\/} of states, but
it remained obscure as what initiates such a decay. 

Further progress was stalled until the evolution of quantum field theory 
in the 1940's. It became apparent then, that the vacuum we 
intuitively see as a void space is indeed a bubbling, fluctuating sea of 
energy. Among the early contributors to this picture was Hiroomi Umezawa,
who realized that this bubbling may couple to the observable world
\cite{U48}. Since this adds a statistical element to the microscopic 
world, it became obvious that irreversibility would find 
its natural explanation in quantum field theory -- and people started 
to think about a connection of field theory to thermodynamics.

It is the purpose of this paper to present a brief review of the 
modern view on disspative quantum field theory, which has evolved
in the 1990's under Hiroomi Umezawas continuous participation \cite{hu92}.

The paper is organized as follows: In the next two sections
the focus will be on the fundamental aspects of
Hilbert space doubling and the breakdown of the quasi-particle picture.
Section 4 is devoted to fundamental physical effects 
in hot plasmas that are connected to finite temperature quantum field 
theory. In section 5 the picture is extended to non-equilibrium systems,
followed by a description of its physical consequences. For these 
physical examples we chose a toy model of ``quarks'' in a hot gas of
massles bosons, as probably present in relativistic heavy 
ion-collisions \cite{QM} and the early universe \cite{Boy}.
\section{Hilbert space doubling}
One of the early findings of finite temperature field theory is
the connection between temperature and temporal boundary conditions: 
The physical excitations propagating into the future 
are either particles, or they are holes in some sea (like e.g. the vacuum 
sea). From this follows, that the occupied states of such 
a sea must propagate backwards in time -- a combination, which gives rise 
to the Feynman boundary condition for vacuum Green functions.
For a system at finite, i.e., nonzero temperature, the temporal boundary 
conditions for Green functions are the Kubo-Martin-Schwinger 
(KMS) conditions \cite{KMS}. 

The KMS conditions take into account the fact, that at finite temperature 
the states of a system are occupied with a certain probability, hence 
with this probability are propagating backwards in time. This 
immediately raises the question, whether one cannot find a new linear 
combination of particles and holes which eliminates this probabilistic 
factor from the description.

For non-equilibrium systems this question is certainly harder to answer 
than for thermal states, because in the latter the occupation probablity 
for each state is fixed. For thermal as well as for non-equilibrium
state one therefore carries along the time-forward
(retarded) as well as time-backward (advanced) boundary condition.
The two-point function of a fermionic quantum field 
with canonical anti-commutation relation
$\left\{\psi(t,\vec{x}) , \vphantom{\int}
        \psi^\dagger(t,\vec{y})\right\} = 
        \delta^3(\vec{x}-\vec{y})$
then has an additional $2\times 2$ matrix structure.
For the implementation of this matrix structure in quantum field
theory exist two ``flavors''. The oldest is the Schwinger-Keldysh or
closed-time path formalism, CTP \cite{SKF}. However, it has the
disadvantage of containing only a single representation of the
canonical anti-commutation relation given above.

From the more fundamental viewpoint one is required to have two such
representations in a thermal system, which are mutually anti-commuting
(commuting for bosons). A proper method to construct these is called
thermo field dynamics (TFD), discovered by Hiroomi Umezawa in 1975 \cite{LMU74,TFD}.

In thermo field dynamics (TFD) the field $\psi$ 
is complemented by a second field
$\widetilde{\psi}_x$ with canonical anti-commutation relations
$
\left\{\widetilde{\psi}(t,\vec{x}) , \vphantom{\int}
\widetilde{\psi}^\dagger(t,\vec{y})\right\}= 
        \delta^3(\vec{x}-\vec{y})
$
and anti-commuting with $\psi$. While $\psi$ is evolving forward in
time, $\widetilde{\psi}$ is subject to a reversed time evolution.
 
For the purpose of this mini-review 
we will treat the two methods CTP and TFD as equivalent because the 
matrix valued propagator of TFD,
\begin{equation}\label{pdf}
S^{(ab)}(x,y)  = 
-\i\left( {
   \array{ll}
    \left\langle \mbox{T}\left[\vphantom{\int}\psi_x 
        \overline{\psi}_y\right]\right\rangle &
    -\i\left\langle\mbox{T}\left[\vphantom{\int}\psi_x 
        \widetilde{\psi}_y\right]\right\rangle
     \gamma^0\\[2mm]
    \i\gamma^0
    \left\langle \mbox{T}\left[\vphantom{\int}\widetilde{\overline{\psi}}_x 
     \overline{\psi}_y\right]\right\rangle &
     -\gamma^0
     \left\langle\mbox{T}\left[\vphantom{\int}\widetilde{\overline{\psi}}_x
\widetilde{\psi}_y\right]\right\rangle
    \gamma^0
   \endarray} \right)
\;,\end{equation}
is exactly equal to the Schwinger-Keldysh result.
For a detailed discussion of this 
equivalence and the exact meaning of $\left\langle\cdot\right\rangle$
we refer to \cite{hu92,h94rep}. One virtue of using the closed-time path 
formalism and TFD together is the easy recognition that {\em by 
construction\/} the above propagators fulfill 
\begin{equation}\label{sme}
S^{11}(x,y)+S^{22}(x,y)=S^{12}(x,y)+S^{21}(x,y)
\;\end{equation}
for equilibrium as well as for non-equilibrium states.

Actually, for the case of thermal equilibrium, 
the KMS boundary condition {\em imposes\/} the
double structure on the Hilbert space of a quantum field theory 
\cite{LW87}. For the above propagator, since in equilibrium it depends only
on $(x-y)$, this KMS condition 
is most easily expressed after a Fourier transform 
$(x-y)\rightarrow P=(p_0,\vec{p})$:
\begin{equation}\label{kmf}
\left(1 - n_F(p_0)\right)S^{12}(p_0,\vec{p}) +
 n_F(p_0) S^{21}(p_0,\vec{p}) = 0
\;,\end{equation}
where 
\begin{equation}
\label{fdir}
n_F(E) = \frac{1}{\displaystyle{\mathrm e}^{\beta(E-\mu)}+1}
\;\end{equation}
is the Fermi-Dirac distribution function at temperature $1/\beta$ and 
chemical potential $\mu$. Because there are now two 
independent linear relationships,
(\ref{sme}) and (\ref{kmf}), among the matrix elements of the 
propagator, it is clear that at least in thermal equilibrium it contains
a lot of spurious information.
\section{Non-shell quantum fields}
A second ingredient, labeled the breakdown of the quasi-particle picture,
is equally important for 
quantum field theory of thermal and non-equilibrium states. 
It is due to the fact, that in general one does not consider systems 
which at (temporal and spatial) infinity consist of free particles. Rather,
the physical systems we are interested in lack such a free asymptotic 
condition and need to be described with a more general asymptotic condition
\cite{L65,L88}. To express this formally, perform the Fourier 
transform of the propagator with respect to the difference $(x-y)$
also in non-equilibrium states, which then defines a mixed (Wigner) 
representation in terms of the momentum variable $P=(p_0,\vec{p})$ 
and $X=(x+y)/2$.

In this representation, the retarded and advanced propagator are 
determined by a dispersion integral
\begin{eqnarray}
\label{split}
\nonumber
S^{R,A}(X;p_0,\vec{p}) &\equiv& S^{R,A}_{XP}=
G_{XP} \mp {\mathrm i}\pi {\cal A}_{XP} \\
  &=& \int\limits_{-\infty}^\infty\!\!dE\;
        \frac{{\cal A}(X;E,\vec{p})}{p_0-E\pm{\mathrm i}\epsilon} 
\;\end{eqnarray}
over a generalized spectral function ${\cal A}_{XP}$, i.e., retarded and 
advanced propagator have a common analytical continuation away from the 
real energy axis. For the free case this function would be
$
{\cal A}(E,\vec{p}) \longrightarrow
        \left( E\gamma^0 + \vec{p}\vec{\gamma} + m\right)\,
        \mbox{sign}(E)\,\delta(E^2-\vec{p}^2 - m^2)
$.
The properties of the ''spectral'' function 
in thermal and non-equilibrium states 
follow from the absence of the free asymptotic 
condition: For thermal systems it has been rigorously proven
\cite{NRT83}, that ${\cal A}$ must not contain isolated poles on the real 
energy axis, i.e., quasi-particles do not exist at finite temperature.
In other words, the irreducible representations of the space-time symmetry
group at finite temperature do not have a mass shell.
Since the dispersion integral furthermore 
shifts {\em complex\/} poles onto the 
unphysical Riemann sheet, the only non-analyticity for 
a thermal or non-equilibrium propagator therefore are cuts along the 
real energy axis. 

In the thermal equilibrium case, the mixed representation has no 
$X$-dependence. One may therefore
use the KMS condition and the linear relation (\ref{sme}) together 
with the definition of retarded and advanced propagator 
to obtain the matrix valued propagator in equilibrium as
\cite{hu92,h94rep}
\begin{eqnarray}\nonumber
&&S^{(ab)}(p_0,\vec{p}) =
   \int\limits_{-\infty}^\infty\!\!dE\,
        {\cal A}(E,\vec{p})\;\times \\ \label{fsk1}
&&\tau_3\, ({\cal B}(n_F(E)))^{-1}\;
   \left(\!{\array{ll}
         {\displaystyle \frac{1}{p_0-E+{\mathrm i}\epsilon}} & \\
    &    {\displaystyle \frac{1}{p_0-E-{\mathrm i}\epsilon}}
\endarray}\right)\;
 {\cal B}(n_F(E))
\;\end{eqnarray}
with a $2\times 2$ Bogoliubov matrix ${\cal B}$
depending on a single parameter $n$ as
\begin{equation}\label{lc}
\label{bdef}
{\cal B}(n) =
\left(\array{lr}(1 - n) &\; -n\\
                1     & 1\endarray\right)
\;.\end{equation}
Diagram rules, transport theory and other developments
of traditional thermal field theory may be carried over to this
treatment in terms of spectral functions \cite{h94rep}.

For the purpose of the present paper it is sufficient to restrict the 
discussion to a simplified ansatz for such a spectral function,
which contains two
independent parameters: A dynamical mass $m$ and a spectral width 
$\gamma$, both are for the sake of a simple approximation taken as
energy and momentum independent
\begin{equation}
{\cal A}(E,\bbox{p}) = \frac{\gamma}{\pi} 
\frac{\gamma^0\left(E^2 + \omega^2+ \gamma^2\right) +
      2 E \bbox{\gamma}\bbox{p} + 2 E m}{
  \left(E^2 - \omega^2 - \gamma^2\right)^2 + 4 E^2 \gamma^2}
\label{af}
\;,\end{equation}
with $\omega^2=m^2+\bbox{p}^2$.
One may regard such a spectral function as the generalization of the
standard $\delta$-function energy-momentum relation to a broader 
distribution for thermally scattered particles, e.g. to a kind of
Lorentzian curve. 
\section{A Physical Effect in Thermal Systems} 
In the following a physical effect is discussed which arises from the
breakdown of the quasi-particle picture in finite temperature field 
theory: If one performs a quasi-particle calculation of the photon 
radiation rate out of a hot plasma, it diverges in the infrared 
(soft photon) sector \cite{kob}. 
It will be shown, how the introduction of a nontrivial 
fermion spectral function cures this problem.
For brevity, we present only a few results, and 
refer to \cite{qh95gam} for the technical details.

In a hot equilibrated plasma, the Hilbert space doubling has to be 
carried out also for the photons, i.e., photon self energies a well as 
propagators exhibit the $2\times 2$ matrix structure also found above.
The {\em radiation rate\/} of these 
photons, i.e., the emission rate through an artificial boundary put into the 
system is given by the unordered (Wightman, or 12-) 
matrix element of the photon self energy $\Pi$.    

Apart from the $2\times 2$ matrix structure due to the Hilbert space 
doubling however, this self energy function also is a tensor in 
Minkowski space. A gauge invariant photon production rate is obtained 
in the sum over all polarizations,
$\epsilon_{\mu} \epsilon_{\nu} \Pi^{\mu \nu} = \Pi_{\mu}^{\mu}$.

The question of gauge invariance requires a careful discussion, because 
usually it necessitates a calculation of $\Pi$ in two-loop order 
(otherwise $\Pi$ will violate current conservation). However, with our 
ansatz spectral function we are on a safe side: The {\em necessary\/}
vertex correction drops out of this sum over polarizations, and one
may obtain a gauge invariant photon radiation rate 
\begin{equation}
 R(E_{\gamma},T) = E_{\gamma} \frac{ dN_{\gamma}}{d^3\vec{p}} = 
 2\frac{n_B(E_\gamma,T)}{8 \pi^3}\,
 \mbox{Im}\left(\Pi^R_{11}+\Pi^R_{22}\right)
=\frac{{\mathrm i}}{8 \pi^3}\,\left(\Pi^{12}_{11} + \Pi^{12}_{22}\right)
\;.\label{prate}
\end{equation}
Here, the lower indices are Lorentz indices, and the upper indices refer
to the $2\times 2$ matrix structure of the Hilbert space doubling. 
$n_B$ is the Bose-Einstein distribution function for the photons, which 
appears when exploiting the KMS condition relating 
the 12 matrix element of $\Pi$ to the imaginary part of 
the retarded polarization function. The latter we take from 
ref. \cite{h94rep} as
\begin{eqnarray}
{\mathrm Im} \Pi^R_{\mu\nu}(k_0,\vec{k}) &  = &
 -\pi\, e^2\,\int\!\!\frac{d^3\vec{p}}{(2\pi)^3}\,
   \int\limits_{-\infty}^\infty\!\!dE 
\label{eqimpi}\\
\nonumber
&&     \mbox{Tr}\left[\gamma_{\mu} {\cal A}(E+k_0,\vec{p}+\vec{k})
    \gamma_{\nu} {\cal A}(E,\vec{p}) \right]\,\left
(n_F(E)-n_F(E+k_0)\right)
\;.\end{eqnarray}
where $e$ is the electric charge of the fermion and ${\cal A}$ is the 
spectral function including the spectral width parameter $\gamma$.

In the next step, a model for the mass and spectral width of the 
fermions is needed. In ref. \cite{qh95gam} it is discussed, how they may be 
obtained for ``quarks'' in a strongly coupled Nambu--Jona-Lasinio model. In 
this model one obtains a strongly temperature dependent ``quark'' mass and 
spectral width due to a chiral phase transition -- and consequently also 
a very interesting temperature dependence of the photon radiation rate. 
Since the present paper is devoted to the fundamentals of dissipative 
field theory, this model is not discussed in greater detail.  

Instead, we have simply plotted the result for the photon radiation rate
for a degenerate plasma of constituent
``quarks'' with 300 MeV mass, and a purely 
electromagnetic spectral width $\gamma^{\mbox{\small em}}$ .
Although this width $\gamma^{\mbox{\small em}}$ is, in principle,
a non-analytical function of the temperature,
the smallness of the electromagnetic coupling constant 
$\alpha=e^2/4\pi$ allows to 
approximate it very well by the lowest order result 
\begin{equation} \label{gamelm}
 \gamma^{\mbox{\small em}}(T) \approx 
 \frac{5}{9}\,\alpha \, T
\;,\end{equation}
where the factor 
$5/9$ is due to the (u,d)-family averaging of the ``quark'' 
electric charge.

\begin{figure}[t]
\vspace*{110mm}
\includegraphics{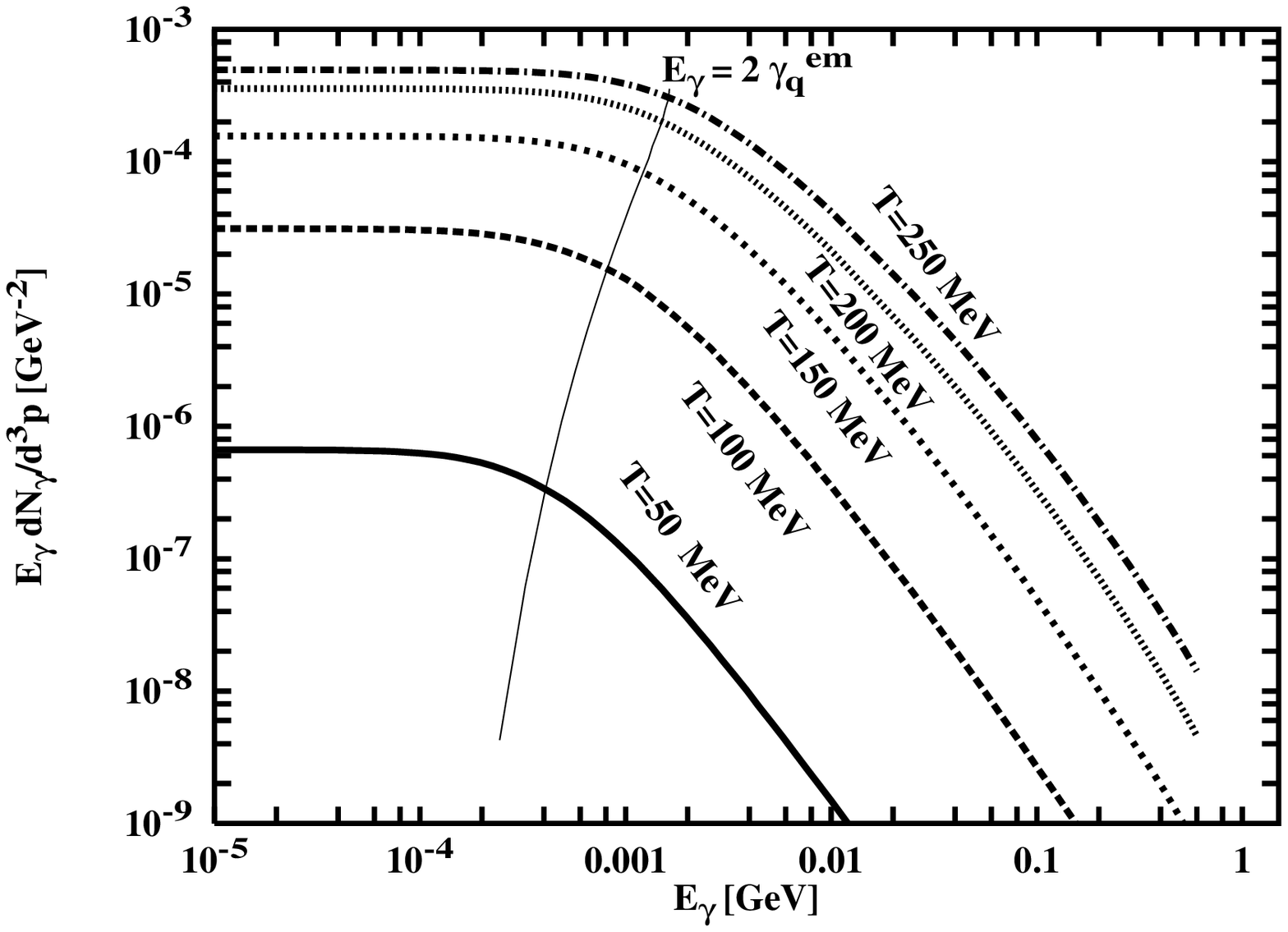}
\caption{Photon production rate $R_{\gamma}$ from an electromagnetically 
interacting particle of 300 MeV mass in a plasma as a function of the 
photon energy $E_{\gamma}$ for different temperatures $T$.}
\label{relm}
\hrule
\end{figure}

In figure \ref{relm} it is shown, that this calculation 
leads to a photon production 
rate which for high photon energies is suppressed by a Boltzmann factor -- 
whereas it saturates for photon energies 
$E_\gamma\le2\gamma^{\mbox{\small em}}$ and does {\em not\/} lead to an 
infrared divergence.

Thus it is found, that when properly taking into account the breakdown 
of the quasi-particle picture, i.e., when substituting a spectral 
function that does not exhibit isolated poles, the infrared problem is 
solved.
 
One may conclude the discussion of this equilibrium physical effect by
relating it to an intuitive picture: The spectral broadening of fermions 
in a hot plasma is due to their repeated thermal scattering, i.e., to a 
kind of Brownian motion. The average distance between two such scattering 
events is $\propto 1/(2 \gamma)$, and consequently the fermion forms an 
''antenna'' for electromagnetic radiation which is not longer than this
distance. This leads to the cutoff of the soft photon radiation.
\section{Relaxation in Non-Equilibrium Systems}
Having outlined a physical effect of the spectral broadening in thermal 
systems, we now turn to the question of dissipation in non-equilibrium 
systems. The results discussed above already indicate, that the spectral 
function parameter $\gamma$, i.e., the spectral width of the 
''particle'', is due to collisions present in a system. Consequently 
one may expect this parameter also to be associated to relaxation processes 
in  non-equilibrium systems.

Such relaxation processesare usually described by transport equations -- 
and hence the task for the present paper is to give an overview of the 
connection between those transport equations on one side and dissipative 
field theory on the other side. To this end, one has to study 
the Schwinger-Dyson equation for the full fermion propagator in coordinate 
space
$
S =  S_0  +  S_0 \odot \Sigma \odot S
$, where
$S_0$ is the free and $S$ the full two-point Green function
of the fermion field, $\Sigma$ is the full self energy
and the generalized product $\odot$ is 
a matrix product (thermal and spinor indices) and an
integration (each of the matrices is a function of two space
coordinates). When switching to the mixed (Wigner-) representation as
introduced above, one has to perform a nontrivial step to handle the 
convolution integrals.  Formally, their Wigner transform may
be expressed as a gradient expansion
\begin{equation}\label{gex}
 \int\!\!d^4(x-y) \;
  \exp\left({\mathrm i} P_\mu (x-y)^\mu\right)\; \Sigma_{xz}\odot G_{zy}
 = \exp\left(-{\mathrm i}\Diamond\right)\,\tilde\Sigma_{XP} \,
  \tilde{G}_{XP}
\;.\end{equation}
$\Diamond$ is a 2nd order differential operator acting on both
functions behind it like a Poisson bracket
$
\Diamond A_{XP} B_{XP}    =  
  \frac{1}{2}\left(\partial_X A_{XP} \partial_P B_{XP}-
                            \partial_P A_{XP}\partial_X B_{XP}\right)
$ 
.
Henceforth will be used the infinite-order differential operator
$
\exp(-{\mathrm i} \Diamond)=
\cos\Diamond-{\mathrm i}\sin\Diamond
$.
Similar to the propagator in eq. (\ref{split}), the self energy is
split into real Dirac matrix valued functions
$\Sigma^{R,A}_{XP} =
  \mbox{Re}\Sigma_{XP} \mp {\mathrm i}\pi \Gamma_{XP}
$.
One then inserts these expressions into the Schwinger-Dyson equation, and 
performs a careful split into real and imaginary parts \cite{h94rep}.
The resulting diagonal components of the matrix valued 
Schwinger-Dyson equation are
\begin{eqnarray}
\label{k8c}
&&\mbox{Tr}\left[\left(
   P^\mu\gamma_\mu- m  \right)  {\cal A}_{XP}\right] =
  \cos\Diamond\,\mbox{Tr}\left[ 
  \mbox{Re}\Sigma_{XP}\, {\cal A}_{XP}
                         + \Gamma_{XP} \, G_{XP}\right]\\ 
\nonumber
&&\mbox{Tr}\left[\left(
   P^\mu\gamma_\mu- m   \right)  G_{XP}\right] =
  \mbox{Tr}\left[1\right] + 
  \cos\Diamond\, \mbox{Tr}\left[ 
  \mbox{Re}\Sigma_{XP} \, G_{XP}
                         -\pi^2\,\Gamma_{XP}\,{\cal A}_{XP}\right]
\;,\end{eqnarray}
i.e., a closed set of differential equations.
Two important facts about these equations have to be emphasized. 
First notice that these equations do not in general 
admit a $\delta$-function solution
for ${\cal A}_{XP}$ even in zero order of $\Diamond$.

Secondly, the equations do not contain odd powers of
the differential operator $\Diamond$. This implies, that when truncating 
the Schwinger-Dyson equation to first order in $\Diamond$
(the usual order for the approximations leading to 
{\em kinetic\/} equations), the spectral function ${\cal A}_{XP}$ 
may still be obtained as the solution of an algebraic equation.

The original Schwinger-Dyson equation was matrix valued. Although
eq. (\ref{sme}) implies, that one of the components is spurious,
a third differential equation still remains to be solved.
It may be transformed into
\begin{eqnarray}\label{tpe1} \nonumber
\mbox{Tr}\left[\left(
  \partial_X^\mu\gamma_\mu + 2 \sin\Diamond\;
  \mbox{Re}\Sigma_{XP}
   + \cos\Diamond\;2\pi\Gamma_{XP}
  \right)  S^K_{XP}\right]& = \\
   2{\mathrm i} \mbox{Tr}\left[
             {\mathrm i}\sin\Diamond\;\Sigma^K_{XP} \, G_{XP}
             - \cos\Diamond\;\Sigma^K_{XP} \, 
    {\mathrm i}\pi{\cal A}_{XP}\right]&
\;.\end{eqnarray}
with $S^K=(S^{12}+S^{21})/2$ and $\Sigma^K=(\Sigma^{12}+\Sigma^{21})/2$.
Note, that here even as well as odd powers of the operator
$\Diamond$ occur, and the solution in zero order $\Diamond$ is
not trivial. To see this more clearly, we {\em define\/} the generalized
covariant distribution function $N_{XP}$ through the equation
\begin{equation}\label{nde}
\left(1-N_{XP}\right)\,S_{XP}^{12} + N_{XP}\,S_{XP}^{21} =0
\,,\end{equation}  
and then find that eq. (\ref{tpe1}) is a differential equation
for $N_{XP}$. Consequently, this third ''off-diagonal'' equation is the
{\em transport equation\/} giving us the desired connection to classical 
physics.

Before drawing some general conclusions, an application of the formalism 
developed here will be discussed briefly. The following toy model is used:
A gas of bosons (gluons) is 
instantaneously heated to a very high temperature. In this 
gas then eventually ``quark-antiquark'' pairs start to pop up,
until at the very end a thermal equilibrium is reached. If these ``gluons'' 
dominate the self energy function for the few ``quarks'' in the medium, and 
if the back-reaction of quarks on the gluons may be neglected, this 
amounts to an imaginary part of the self energy function as
\begin{equation}\label{ss1}
\Gamma_{XP}\equiv\Gamma_t = \gamma^0\;g T(t)\,= \gamma^0\;g\,
\left( T_i\,\Theta(-t)\,+\,T_f\,\Theta(t) \right)
\;\end{equation}
Furthermore, one may  neglect the influence of anti-quarks in 
the spectral function (\ref{af}), but now the dynamical mass (and
therefore $\omega\equiv\omega_t$) and spectral width
$\gamma\equiv\gamma_t$ are time-dependent.

With this spectral function, the coupled system (\ref{k8c})  
reduces to {\em a single\/} nonlinear equation
for $\gamma_t$ plus the condition 
$
\omega^2_t = \omega^2_0 = \vec{p}^2 + m^2
$. 
This latter condition is more complicated, when the anti-particle piece of 
the spectral function is taken into account.
The energy parameter is chosen as $E=\omega_0$, which yields instead of
eq. (\ref{k8c}) as the Schwinger-Dyson equation for the retarded
(or advanced) two-point function of the quarks:
\begin{equation}\label{k9c}
\gamma_t = g T_i + g (T_f - T_i)\,\Theta(t)\,
   \left(1-{\mathrm e}^{{\displaystyle -2 \gamma_t t}}\right)
\end{equation}
In fig. 2, the solution of this equations is plotted in comparison
to the time dependent imaginary part of the self energy function
from eq. (\ref{ss1}). It is obvious, that the solution of the
nonlinear equation (\ref{k9c}) approaches
the imaginary part of the self energy function with a 
characteristic delay time. 

Now consider two different levels of transport theory
for this model, the corresponding generalized distribution functions
are labeled $N_t$ and  $N^B_t$.
First of all, due to the simplicity of
our spectral and self energy function ansatz,
the full quantum transport equation (\ref{tpe1}) reduces to 
\begin{equation}\label{tpe2}
\frac{d}{d t}N_t = -2 \,\gamma_t\left( N_t - n_F(m,T(t)) \right)
\;\end{equation}
with $T(t)$ as defined in eq. (\ref{ss1}). Indeed it turns out, that the 
spectral width parameter is responsible for the irreversible time 
evolution of the non-equilibrium state.

This equation looks surprisingly similar to a kinetic equation in
relaxation time approach. However, this similarity is superficial:
The {\em kinetic\/} equation,
or Boltzmann equation, derived for our simple model system reads
\begin{equation}\label{tpe3}
\frac{d}{d t}N^B_t = -2 \,\Gamma_t\left( N^B_t - n_F(m,T(t)) \right)
\;.\end{equation}
The difference between the two transport equations in this simple toy 
model is therefore the occurence of the self-consistent spectral 
width parameter in the quantum transport equation, whereas the Boltzmann
relaxation parameter is given by the imaginary part of the self energy.

In fig.3 the numerical solution for $N_t$ is shown and compared
to the Boltzmann solution $N^B_t$.
The comparison of the two methods shows, 
that the full quantum transport equation
results in a {\em much \/} slower equilibration process
than the Boltzmann equation.

This result is in agreement with other attempts to solve
the quantum relaxation problem: The quantum system
exhibits a memory, it behaves in an essentially non-Markovian way.
The reason for this behaviour is the time-dependent spectral width, which 
follows the imaginary part of the self energy function only with some 
delay time.

In particular, for the physical scenario studied here,
the time to reach 1-1/e${}^2\approx$ 86 \% of the equilibrium 
``quark'' occupation number is almost doubled 
(14.7 fm/c as compared to 8.2 fm/c in the 
Boltzmann case). 

Thus, although we only used a toy model, 
it might turn out that quantum effects (= memory as described in this
contribution) substantially hinder the thermalization of a 
strongly interacting plasma over long time scales. 
A more thorough discussion of this physical result is carried out
in ref. \cite{hbfz96}.
\section{Conclusion}
In the preceding sections it was shown, how the two principal features 
of dissipative field theory, i.e., {\em 1.Hilbert space doubling\/} and
{\em 2.Continuous spectral functions\/} provide a unified description
of equilibrium and non-equilibrium phenomena in statistical systems.

In particular, the spectral width parameter provides the regularization of
unphysical divergences as well as the relaxation rate for non-equilibrium 
states; it may be considered a system parameter as important as the 
dynamical mass. Of course, the toy model examples we have considered here 
are much too simple for deeper physical conclusions -- in a realistic 
system, the energy-momentum dependence of the spectral function is 
certainly not negligible.

It is also obvious by now, that indeed the dissipation we experience in 
the macroscopic world has its expression already on the microscopic level:
In quantum field theory of thermal systems, each state has an infinite
number of infinitely close neighbor states. Thus, while for any finite 
number or even countably infinite number of degrees of freedom the time 
evolution of a system will be reversible, this is no longer the case in 
quantum field theory. Continuous spectral functions, as fundamental 
feature of thermal systems, will always exhibit irreversible behaviour.

Let us as the final part of this paper add a comment on the connection 
of this unified view with the work of Hiroomi Umezawa. As seen above,
the equilibrium propagator (\ref{fsk1}) admits a {\em diagonalization\/}
with a Bogoliubov transformation matrix ${\cal B}$. This concept of a 
thermal Bogoliubov transformation was first introduced by Umezawa and 
collaborators \cite{LMU74}, and finally has led to the diagonalization 
transformation \cite{hu92}. One may now close the arc of the 
argumentation by realizing that the proper {\em 
non-equilibrium\/} solution $N_{XP}$ of the generalized transport 
equation (\ref{tpe1}) allows for a {\em diagonalization\/} also in general 
states \cite{h94rep}:
\begin{equation}
{\cal B}(N_{XP})\,\tau_3\,S_{XP}\,({\cal B}(N_{XP}))^{-1}=
\left({\array{rr} G_{XP}-{\mathrm i}\pi{\cal A}_{XP} & \\
 & G_{XP}+{\mathrm i}\pi{\cal A}_{XP} \endarray}\right)
\;.\end{equation}
The physical and mathematical understanding of this Bogoliubov symmetry 
has been achieved in thermo field dynamics. The {\em diagonalization 
condition\/} in equilibrium states is equivalent to the KMS condition, 
i.e., it specifies the {\em diagonalization parameter\/} unambigously
as a Bose-Einstein or Fermi-Dirac distribution function. In 
non-equilibrium states, the parameter is obtained as the solution of a 
transport equation \cite{hu92}. 

This unified view therefore may be seen 
as the final achievement of Hiroomi Umezawa, and I am very grateful to
the Creator of all worlds that I had the opportunity to share this
achievement with Dr. Umezawa. 

\begin{figure}[t]
\vspace*{75mm}
\includegraphics{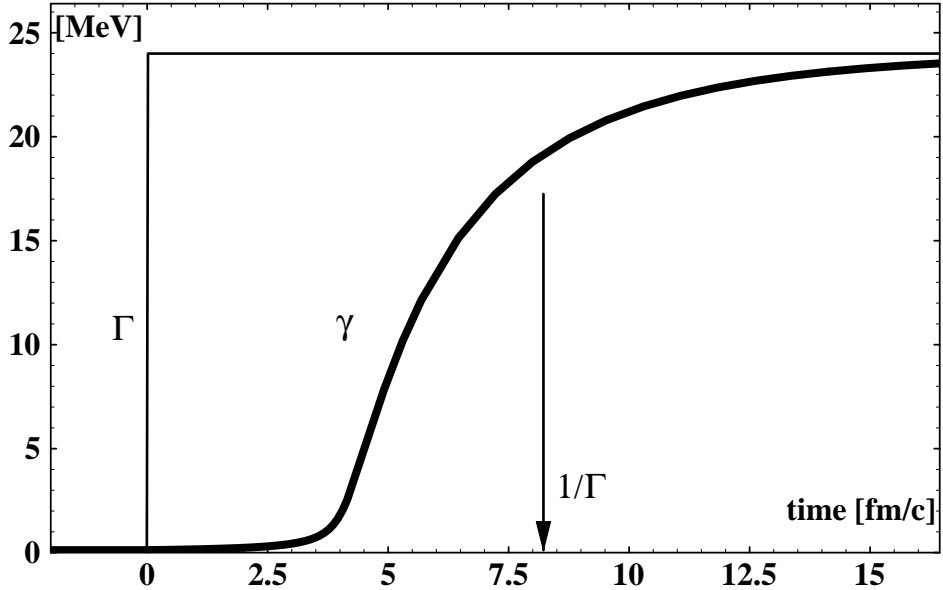}
\caption{Time dependent spectral width parameter
$\gamma_t$.\protect\newline
Parameters are $g$=0.12,  $T_i=$ 1 MeV, $T_f=$ 200 MeV,
$m=$ 10 MeV.\protect\newline
Thin line: $\Gamma_t$ from eq. (\ref{ss1}),
thick line: $\gamma_t$ from eq. (\ref{k9c}).
}
\vspace*{1mm}
\hrule
\end{figure}
\begin{figure}[t]
\vspace*{75mm}
\includegraphics{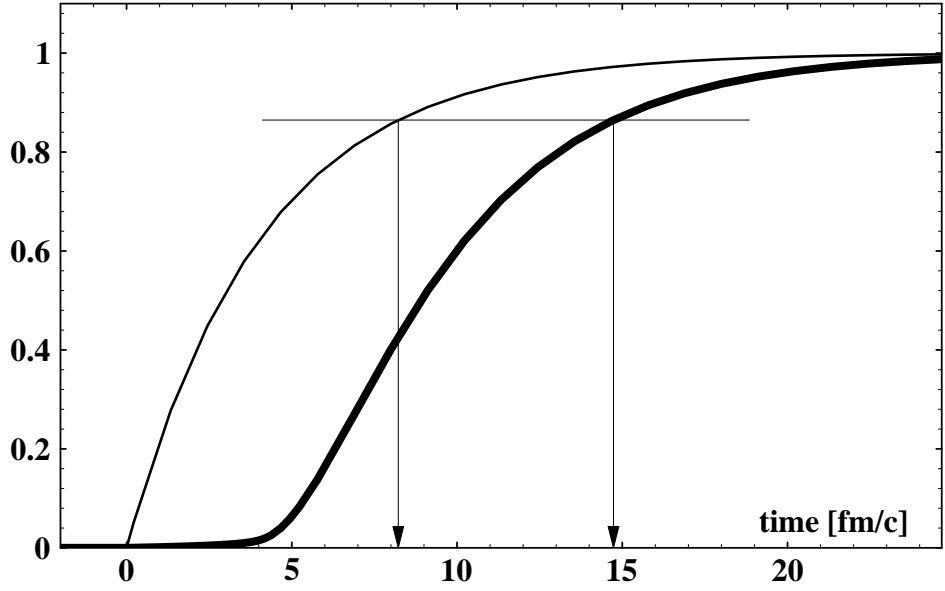}
\caption{Normalized time dependent fermionic distribution function for 
 slow quarks.\protect\newline
Parameters as in Fig. 2;
thin line $N^B_t/n_F(m,T_f)$ from the Boltzmann equation (\ref{tpe3}),
thick line $N_t/n_F(m,T_f)$ from the quantum transport equation (\ref{tpe2});
}
\vspace*{1mm}
\hrule
\end{figure}
\end{document}